\documentstyle[prc,aps,preprint,epsfig]{revtex}
\begin{document}
\draft
\title{Treatment of competition between complete fusion and
quasifission in collisions of heavy nuclei}
\author{ G.G.Adamian$^{1,2}$, N.V.Antonenko$^{1,2}$,
W.Scheid$^{1}$ and V.V.Volkov$^{2}$}
\address{$^{1}$Institut f\"ur Theoretische Physik der
Justus--Liebig--Universit\"at,
D--35392 Giessen, Germany\\
$^{2}$Joint Institute for Nuclear Research, 141980 Dubna, Russia}
\date{\today}
\maketitle

\begin{abstract}
A model of competition between complete fusion and quasifission
channels in fusion of two massive nuclei is extended to
include the influence of dissipative effects
on the dynamics of nuclear fusion.
By using the multidimensional Kramers-type stationary solution
of the Fokker--Planck equation, the fusion rate through the
inner fusion barrier in mass asymmetry is studied.
Fusion probabilities in symmetric
$^{90}$Zr+$^{90}$Zr, $^{100}$Mo+$^{100}$Mo,
$^{110}$Pd+$^{110}$Pd, $^{136}$Xe+$^{136}$Xe, almost symmetric
$^{86}$Kr+$^{136}$Xe and $^{110}$Pd+$^{136}$Xe reactions
are calculated. An estimation of the fusion probabilities
is given for asymmetrical
$^{62}$Ni+$^{208}$Pb, $^{70}$Zn+$^{208}$Pb,
$^{82}$Se+$^{208}$Pb, and
$^{48}$Ca+$^{244}$Pu reactions used for
the synthesis of new superheavy elements.

\end{abstract}


\section{Introduction}
The competition between the complete fusion and quasifission
processes occurs in the reactions with massive nuclei at
bombarding energies smaller than 15 MeV/nucleon.  In these
reactions the quasifission channel dominates and leads to a
strong reduction of few orders of magnitude of the fusion cross
section $\sigma_{CN}$ \cite{1}. The competition between
the complete fusion and quasifission processes was not
considered in the calculations with various models, namely with
the macroscopic dynamical \cite{2}, optical \cite{3} and
surface friction \cite{4} models.  It was shown in  \cite{1}
that these models do not lead to correct fusion cross sections
for reactions with heavy nuclei \cite{5,25,6}. The new model
suggested in \cite{1} yields a good agreement between the
theoretical predictions and experimental data on the fusion of
heavy nuclei.  Within this model the fusion cross section
$\sigma_{CN}$ is defined as
\begin{equation}
\sigma_{CN}(E_{\rm c.m.})=\sum_{J=0}^{J_{max}}\sigma_{c}(E_{\rm c.m.},J)
P_{CN}(E_{\rm c.m.},J).
\label{1_eq}
\end{equation}
Here, $\sigma_{c}$ is the capture cross section at
the bombarding energy $E_{\rm c.m.}$ and
angular quantum number $J$ and $P_{CN}$ the
fusion probability after the capture stage of collision,
which takes the competition between
the fusion and quasifission processes into account.
The value of $J_{max}$ depends on $E_{\rm c.m.}$ and
is smaller than $J_{B_f=0}$ at which the fission barrier
in the compound nucleus vanishes \cite{arm}.
In order to calculate $\sigma_{c}$,
either optical \cite{3} or
surface friction \cite{4} models can be used \cite{1}.
$P_{CN}=1$ is supposed in the models of Refs.~\cite{3,4}.  In
the macroscopic dynamical model \cite{2} $P_{CN}=1$ for
$E_{\rm c.m.} > B_C + E_{\rm xx}$ and
$P_{CN}=0$ for $E_{\rm c.m.} < B_C + E_{\rm xx}$ where
$B_C$ and $E_{\rm xx}$ are the Coulomb
barrier and the extra--extra push energy, respectively. The
main advantage  of the model \cite{1} in contrast to the models
\cite{2,3,4} is the treatment of the competition between the
complete fusion and quasifission in the calculation of $P_{CN}$.
In the present paper the model \cite{1} is extended by
including dissipative effects on the fusion dynamics.

The fusion process is considered in \cite{1} as the evolution
of the dinuclear system (DNS) in which nucleons are transfered
from the light nucleus to the heavy one \cite{7}. The
DNS-concept is based on the information obtained from the
investigation of deep inelastic collisions of heavy ions
\cite{8}. The DNS is formed at the initial stage of the
reaction  when the kinetic energy is transformed into the
excitation energy of the nuclei.  The initial DNS is localized
in the minimum of the pocket of the nucleus-nucleus potential
$V(R)$ (Fig.~\ref{1_fig}) at $R=R_m$ where $R$ is the relative
distance between the interacting nuclei. In our approach we have
$R_m\approx R_1+R_2+0.5$ fm where $R_1$ and
$R_2$ are the radii of the nuclei
in the DNS. Then the DNS evolves by a diffusion process in the
mass asymmetry degree of freedom $\eta=(A_1 - A_2)/A$ to the
compound nucleus and fuses (Fig.~\ref{1_fig}). $A_1$ and $A_2$
are the mass numbers of the nuclei and $A=A_1 + A_2$. Besides
the motion in $\eta$ a diffusion process in the relative
distance occurs. This process leads to the decay of the DNS
which we denote as quasifission.
For quasifission the DNS
should overcome the potential barrier $(B_{qf})$  which
coincides with the depth of the pocket in the potential $V(R)$
(Fig.~\ref{1_fig}). The important peculiarity of the DNS
evolution to the compound nucleus is the appearance of an inner
fusion barrier $B_{fus}^*$ in the mass asymmetry degree of
freedom with its top (the Businaro-Gallone point) at
$\eta=\eta_{BG}$ which coincides with the maximum of the DNS
potential energy as a function of $\eta$ (Fig.~\ref{1_fig}).
The value of $B_{fus}^*$ supplies the hindrance for the complete
fusion in the DNS-concept.  The energy required to overcome the
fusion barrier $B^*_{fus}$ is contained in the DNS excitation
energy.  As was found in \cite{1,9,10}, the value of
$B_{fus}^*$ can be much smaller than the extra-extra push energy
predicted in \cite{2}.

The dissipative large-amplitude collective nuclear motions,
which occur in fission, quasifission, fusion and heavy-ion
reactions, can be analyzed within the transport theory.  It is
known that the  Kramers expression \cite{11}  yields an
excellent approximation to the fission rate in comparison to the
solution of the Fokker-Planck and Langevin equations
\cite{12,17,13}.  The fusion probability $P_{CN}$ (\ref{1_eq})
can be obtained with the Kramers-type expression for the rate of
fusion through the inner fusion barrier.  The main advantage of
the Kramers-type expression for the fusion rate in contrast to
the statistical method \cite{1} is the possibility to include
nuclear viscosity in fusion process. In this paper, the
two-dimensional ($R$ and $\eta$) Kramers-type expression is
applied to the calculation of $P_{CN}$. In the following we
analytically demonstrate the dependence of the fusion
probability on the values of the friction coefficients, the
temperature $T$ of the DNS, and the values of $B^*_{fus}$ and
$B_{qf}$.  As examples, the symmetrical, almost  symmetrical,
and asymmetrical reactions will be considered.

\section{Model}
\subsection{Fusion probability}
The variables $R$  and $\eta$  are the relevant collective
variables used in \cite{1,14} to describe the DNS evolution.
The neck degree of freedom \cite{2,15,16} which is important in
macroscopic dynamical models is not considered in this paper.
As follows from our analysis \cite{16}, the neck is not a
relevant collective variable for values of $R>R_1+R_2$ which are
important
in the DNS. The size of the neck in the liquid drop model and the
overlap region in the frozen density approximation
are close to each other for
$R_1+R_2\le R\le R_1+R_2+1\, $fm \cite{16}.
Based on this fact and the time scale,
which is of interest, we assume that the  individuality of
the DNS nuclei in the fusion process is retained, and follow
the approach suggested in \cite{1,10}.

Since the fusion probability increases with the mass asymmetry,
the DNS with larger mass asymmetry supplies favorable conditions
for complete fusion. We assume
that  the fusion occurs inevitably for  $\eta>\eta_{BG}$.
The diffusion in $\eta$ is important in
our consideration of the fusion process.
The simultaneous investigation of the transport processes in
$\eta$ and $R$ variables allows us to calculate the fusion and
quasifission probabilities \cite{10}.
In our opinion, the fusion probability
is small in many reactions with
heavy nuclei because the initial mass asymmetry $\eta_i$
is smaller than $\eta_{BG}$ and the diffusion
in $\eta$ is much smaller than the diffusion in $R$.

The leakage of probability through the fusion barrier in $\eta$
is defined by the rate $\lambda_\eta(t)$ at $\eta=\eta_{BG}$
(Fig.~\ref{1_fig}).  Then we obtain
\begin{eqnarray}
P_{CN}= \int\limits_{0}^{t_0}\lambda_\eta(t)dt
\label{2_eq}
\end{eqnarray}
Here, $t_0$ is the life-time of the DNS with
\begin{eqnarray}
\int\limits_{0}^{t_0}
[\lambda_R(t)+\lambda_\eta(t)]dt=1.
\label{21_eq}
\end{eqnarray}
For symmetrical and almost
symmetrical systems,
the symmetry of the fusion process with
respect to $\eta=0$ should be taken into account.
Then the initial DNS is near the
minimum of the potential energy
$U(\eta)$ which is a symmetric function with respect to $\eta=0$
(Fig.~\ref{1_fig}).
Fusion occurs when DNS reaches the barriers
at either $\eta=\eta_{BG}$ or $\eta=-\eta_{BG}$.
The time dependence of the rates $\lambda_i(t)$ ($i=R,\eta$)
can be taken in the following way
\begin{eqnarray}
\lambda_i(t)= \lambda^{Kr}_i\left(\frac{e^{t/\tau_i}-1}{e-1}
\theta(\tau_i-t) + \theta(t-\tau_i)\right),
\label{22_eq}
\end{eqnarray}
where $\lambda^{Kr}_i$ are asymptotic values of the fusion or
quasifission rate $\lambda_i(t)$ at the corresponding barriers
(Fig.~\ref{1_fig}) and $\theta(t)$ is a step function.
Here, we assume
that after the exponential growth during the
transient time $\tau_i$ the rate $\lambda_i(t)$ reaches
the asymptotic value.

Using (\ref{22_eq}) we obtain from Eqs. (\ref{2_eq}) and (\ref{21_eq})
\begin{eqnarray}
P_{CN}=P_{CN}^0-\Delta P_{CN}=
\frac{\lambda_{\eta}^{Kr}}{\lambda_R^{Kr}+\lambda_\eta^{Kr}}-
\frac{\lambda_{\eta}^{Kr}\lambda_R^{Kr}}{\lambda_R^{Kr}+\lambda_\eta^{Kr}}
\frac{\tau_\eta-\tau_R}{\beta},
\label{23_eq}
\end{eqnarray}
\begin{eqnarray}
t_0=t_{00}+\Delta t=\frac{1}{\lambda_R^{Kr}+\lambda_\eta^{Kr}}+
\frac{\lambda_R^{Kr}\tau_R+\lambda_\eta^{Kr}\tau_\eta}
{(\lambda_R^{Kr}+\lambda_\eta^{Kr})\beta},
\label{24_eq}
\end{eqnarray}
where $\beta=e-1\approx 1.72$.
Assuming instead of (\ref{22_eq}) the linear grows of $\lambda_i(t)$
\cite{111}
\begin{eqnarray}
\lambda_i(t)= \lambda^{Kr}_i\left(\frac{t}{\tau_i}
\theta(\tau_i-t) + \theta(t-\tau_i)\right)
\label{25_eq}
\end{eqnarray}
we obtain the Eqs. (\ref{23_eq}) and (\ref{24_eq}) but with
$\beta=2$.
The first terms in (\ref{23_eq}) and (\ref{24_eq})
are the contributions from the quasistationary rates.
The second terms are related to the transient time.
One can see from (\ref{23_eq}) that we can neglect the transient
time for $\tau_i\ll 1/\lambda_i^{Kr}$ ($i=R,\eta$) or
$\tau_R\approx\tau_\eta$.
This is true for all reactions under consideration excepting
the $^{136}$Xe+$^{136}$Xe and $^{110}$Pd+$^{136}$Xe reactions
where the differences between $B^*_{fus}$ and $B_{qf}$ are very large.
Note that the role of the transient stage decreases with
decreasing excitation energy of the DNS because the exponential
increase of $1/\lambda_R^{Kr}$ is larger than the logarithmic increase
of the transient times (see equations below).

To obtain the asymptotic fusion and quasifission rates
$\lambda^{Kr}_i$ ($i=R,\eta$), we use the formalism elaborated
in Refs. ~\cite{12,13}. We approximate the expression for the
quasistationary rate $\lambda^{Kr}_i$ over a multidimensional
potential barrier \cite{13} with a Kramers-type formula
\begin{eqnarray}
\lambda ^{Kr}_{i}=\frac{1}{2\pi}
\frac{\omega^2_i}
{\sqrt{\omega^{B_R}_i\omega^{B_\eta}_i}}
\left(\sqrt{\left[\frac{(\Gamma/\hbar)^2}
{\omega^{B_R}_i\omega^{B_\eta}_i}\right]^2+4}
-\frac{(\Gamma/\hbar)^2}
{\omega^{B_R}_i\omega^{B_\eta}_i}\right)^{1/2}
\exp\left[-\frac{B_i}{T}\right].
\label{5_eq}
\end{eqnarray}
Here, $B_i$ ($i=R,\eta$) defines the height of the fusion
($B_\eta=B^*_{fus}$) or quasifission ($B_R=B_{qf}$) barriers.
The possibility to apply the Kramers-type expression to
relatively small barriers ($B_i/T > 0.5$) was demonstrated in
\cite{17}. The local thermodynamic temperature $T$ is
calculated with the expression $T=\sqrt{E^*/a}$, where $a=A/12$
MeV$^{-1}$ and  $E^*$ is the DNS excitation energy.  In Eq.
(\ref{5_eq}), the frequencies $\omega^{B_j}_i$ ($j=R,\eta$) of
the inverted harmonic oscillators approximate the potential in
the variables $i=R,\eta$ on the tops of the barriers $B_j$, and
$\omega_i$ are the frequencies of the harmonic  oscillators
approximating the potential in the same variables for  the
initial DNS. Since the local oscillator approximation of the potential
energy surface is good for the reaction considered, we neglected
the nondiagonal components of the curvature tensors in
(\ref{5_eq}).  In our calculations, we use the simple
approximate expressions for the friction coefficients
$$\gamma_{ii'}=\Gamma\mu_{ii'}/\hbar,\quad  (i,i'=R,\eta),$$
which were obtained by the linear response theory \cite{18}.
The quantity $\Gamma$ denotes an average double width of the
single-particle states.
The calculation of the mass  parameters
$\mu_{RR}$ and $\mu_{\eta\eta}$ is given in \cite{10,19}.
The role of the nondiagonal components of the tensors
of inertia and friction depends very much on the choice of the
collective variables \cite{18,19}. By using the collective variables
$R$ and $\eta$ for describing the evolution of the DNS
we neglect the nondiagonal component of the tensor of inertia
because $\mu_{R\eta}\ll\sqrt{\mu_{RR}\mu_{\eta\eta}}$
for $|\eta|<|\eta_{BG}|$ \cite{19}.
The nondiagonal mass coefficient
$\mu_{R\eta}=0$ in the DNS takes an essential role only
for $|\eta|>|\eta_{BG}|$.
As was shown in \cite{10,16}, the friction coefficients
$\gamma_{RR}$ and $\gamma_{\eta\eta}$ obtained
with $\Gamma=2$ MeV have the same order of magnitude as
the ones calculated within the other approaches.

\subsection{Potential energy of the DNS}
The value of
$$\omega^{B_j}_i=\sqrt{\left|\frac{\partial^2U(R, \eta, J)}
{\partial i^2}\right|_{B_j}/\mu_{ii}}$$
is easily calculated with the potential energy of the DNS
\begin{eqnarray}
U(R, \eta, J)&= &B_1+ B_2+  V(R, J)- [B_{12}+V^{'}_{rot}(J)].
\label{7_eq}
\end{eqnarray}
Here, $B_1$, $B_2$, and $B_{12}$ are the binding energies
of the fragments and the compound
nucleus and are calculated with
liquid-drop masses for large excitation energies
and with realistic masses \cite{20} for
small excitation energies.
The isotopic composition of the nuclei
forming the DNS is chosen with the condition
of a $N/Z$-equilibrium in the
system. The value of $U(R, \eta, J)$ is normalized to
the energy of the rotating compound nucleus by
$B_{12}+V^{'}_{rot}$.
The nucleus-nucleus potential $V(R,J)$ in (\ref{7_eq})
is calculated as described in \cite{21}.
The retaining individuality of the DNS nuclei during the time
which is of interest allows us
to calculate the DNS potential energy by the method presented
in \cite{21}.
The calculated driving potential $U(R_m, \eta,J=0)=U(\eta)$
as a function of $\eta$ and the nucleus--nucleus
potential $V(R,J=0)$ as
a function of $R$ in the reaction
$^{90}$Zr + $^{90}$Zr is presented in Fig.~\ref{1_fig}.

The potential energy of the DNS as a function of $\eta$ and
$R$ depends on the temperature, shell effects and angular momentum.
In the present paper we do not analyse in
details the dependence
of the potential energy surface on temperature
because this demands the introduction
of an additional parameter. We distinguish two cases: The first
case corresponds to large excitation energies of the initial DNS
when the liquid drop binding
energies and spherical shapes of the nuclei in the DNS can be used
in the calculations.
The second case corresponds to the cold fusion with
small $E^*$ when
the realistic binding energies are taken in (\ref{7_eq}).
For small values of $E^*$, the deformation of the nuclei in their
ground states \cite{rom} is taken in the DNS into account when
the barrier heights are calculated.
In order to demonstrate
the influence of the shell and deformation effects at small $E^*$,
the calculation of $U(R_m,\eta,J=0)$ is presented
in Fig.~\ref{11_fig} for the $^{86}$Kr+$^{136}$Xe reaction.
In this reaction the deformation of the nuclei in the DNS
leads to a decrease of $B^*_{fus}$.
After the smoothing over
small oscillations caused by even-odd effects we can use
the expression (\ref{5_eq}) for $\lambda_i^{Kr}$
in first approximation.
The calculated driving potentials for the reactions leading
to superheavy nuclei are presented in \cite{ant}.
Below we discuss the case of
the $^{86}$Kr+$^{136}$Xe reaction where
the fusion probability is calculated with driving potentials
based on spherical and deformed shapes of the nuclei.
The calculations of complete fusion probabilities
with spherical (the liquid drop binding energies)
and deformed (the realistic binding energies) nuclei
give us the interval where realistic
values of $P_{CN}$ are situated.
As noted in Ref.~\cite{10}, the values
of $P_{CN}$ for small $E^*$ can be smaller or larger than
the ones for large $E^*$.
In our calculations of the fusion probability, the following values
$\hbar\omega_R^{B_R}\approx 0.8-1$ MeV,
$\hbar\omega_R^{B_\eta}\approx 3-3.5$ MeV,
$\hbar\omega_\eta^{B_R}\approx 1-1.5$ MeV,
$\hbar\omega_\eta^{B_\eta}\approx 1.5-2$ MeV,
$\hbar\omega_R\approx 1.5-2$ MeV, and
$\hbar\omega_\eta\approx 0.8-1$ MeV are used
for the reactions considered.

The dependences of the barrier heights
on $J$ for the reactions $^{90}$Zr+$^{90}$Zr and
$^{110}$Pd+$^{110}$Pd are presented in Fig.~\ref{12_fig}.
The values of $B^*_{fus}$ and $B_{qf}$ are slightly changed
when $J$ increases from 0 to 25$\hbar$.
For heavier systems, this changes are certainly smaller because
of the larger moment of inertia.
As a result,
the values of $P_{CN}$ slightly differ from
those calculated with $J=0$.
In order to calculate $\sigma_c(E_{\rm c.m.},J)$
in (\ref{1_eq})  the simple expression
$\sigma_c(E_{\rm c.m.},J)=\pi\lambdabar^2(2J+1)T(E_{\rm c.m.},J)$
is often used ($\lambdabar$ is the reduced de Broglie wavelength).
The transmission coefficient $T(E_{\rm c.m.},J)$
through the Coulomb barrier restricts also the angular momentum range.
The weight function $(2J+1)P_{CN}(E_{\rm c.m.},J)$ used
to obtain $\sigma_{CN}$ has its maximum at $J=20\hbar$.
Since the calculation of $\sigma_{CN}$ is of interest to
determine the evaporation residues cross sections, only low
angular momenta can be considered. Indeed, the surviving probabilities
of the compound nuclei in the reactions considered are narrow
functions peaking at all energies at $J$ values
in the vicinity of zero \cite{arm}.
Although the precise calculation of the
excitation function demands the dependence of $P_{CN}$ on $J$,
for the estimations of the evaporation residues cross sections,
we can use the values of $\sigma_{CN}$ calculated with
$J_{max}=10-15\hbar$ and $P_{CN}(E_{\rm c.m.},J=0)$
with a good accuracy.

\section{Results and discussions}
\subsection{Effect of transient times}
Since the difference between $B^*_{fus}$ and $B_{qf}$ and their
absolute values
can be large in some reactions, the role of the transient time
in the calculation of $P_{CN}$ and $t_0$ should be estimated.
The motions in $R$ and $\eta$ are close to the
underdamped and overdamped motions, respectively. Therefore,
the transient time for the realistic values of $\Gamma$
can be estimated by the expressions \cite{111}
\begin{eqnarray}
\tau_R=\frac{\hbar}{\Gamma}\ln(10B_{qf}/T),
\label{91_eq}
\end{eqnarray}
\begin{eqnarray}
\tau_\eta=\frac{\Gamma}{2\hbar\omega_\eta^2}\ln(10B^*_{fus}/T),
\label{92_eq}
\end{eqnarray}
The calculated time dependences $\lambda_\eta(t)$ with
Eq.~(24) in \cite{10} are presented in Fig.~\ref{13_fig} for the reactions
$^{90}$Zr+$^{90}$Zr and $^{136}$Xe+$^{136}$Xe at $\Gamma=2$ MeV and
$E^*=30$ MeV. We find that the transient times taken
from Fig.~\ref{13_fig} are practically the same as the ones calculated
with (\ref{92_eq}). The calculated transient times and life-times
for different systems are given in Table~1. One can see that
the transient stage does practically not effect the values of $t_0$.

It is known \cite{pom} that the consideration of the
transient stage is important for large excitation energies
when the particle emission change the system
during the fission or fusion time.
However, this is not so in our cases
and the calculation with the transient time (Table~2)
leads only to a decrease
of $P_{CN}$ by maximal 30\% in the reactions
$^{136}$Xe+$^{136}$Xe and $^{110}$Pd+$^{136}$Xe in comparison with
the value of $P_{CN}^0$ in (\ref{23_eq}) for $\Gamma=2$ MeV.
In other reactions
considered the effect of the transient stage is negligible.
Therefore, we can neglect the transient time in the calculations
of $P_{CN}$ for the most considered reactions.
For the reactions with $B^*_{fus}-B_{qf}\ge 15$ MeV, the values of
$P_{CN}$ are small and the corrections $\Delta P_{CN}$
from the transient stage are practically within the accuracy of
the calculated barrier heights.
In these reactions the values of $P_{CN}$ calculated without
$\Delta P_{CN}$ in (\ref{23_eq}) are the upper limits
of the fusion probabilities and can also be used for
estimations of fusion cross sections.

\subsection{Large excitation energies of the DNS}
The values of $P_{CN}$ (Table~3) are in agreement
for most of the reactions
with the ones extracted from the experimental data
\cite{6,22,23} and with the results of our previous
calculations \cite{1,9}.
For example, for the reactions
$^{90}$Zr+$^{90}$Zr, $^{100}$Mo+$^{100}$Mo, and
$^{110}$Pd+$^{110}$Pd the values of
$P_{CN}\approx 4\times 10^{-1},\ 10^{-2},\ 10^{-4}$, respectively,
yield a good agreement with the experimental
data of $\sigma_{CN}(E_{\rm c.m.})$.
For the $^{110}$Pd+$^{110}$Pd reaction, the value of
$\sigma_{CN}(E_{\rm c.m.})$ calculated
with the macroscopic dynamical model \cite{1}
is about three orders of magnitude larger than the
experimental one. Thus, the competition
between the complete fusion and quasifission processes
is extremely important in the DNS evolution.
Perhaps, due to the small value of $P_{CN}$ in
the $^{110}$Pd+$^{136}$Xe reaction,
the fusion was not observed in \cite{23}.

The values of $P_{CN}$ can be also calculated from (\ref{2_eq})
by using the one--dimensional Kramers-type expression instead of
Eq. (\ref{5_eq}):
\begin{eqnarray}
\lambda ^{Kr}_{i}=\frac{1}{2\pi}
\frac{\omega_i}
{\omega^{B_i}_i}
\left(\sqrt{\left(\frac{\Gamma}{2\hbar}\right)^2+
\left(\omega^{B_i}_i\right)^2}
-\frac{\Gamma}{2\hbar}\right)
\exp\left[-\frac{B_i}{T}\right].
\label{9_eq}
\end{eqnarray}
We approximately find the same results with this formula as with
Eq. (\ref{5_eq}). Therefore, the estimations of the transient
times with (\ref{91_eq}) and (\ref{92_eq}) are realistic.

In addition to the reactions presented in Table~3 the fusion
probabilities were calculated for the reactions $^{96}$Zr+$^{124}$Sn
and $^{124}$Sn+$^{124}$Sn with $\Gamma=2$ MeV, $J=0$ and
the DNS excitation energy 30 MeV. The values of $P_{CN}$ as
a function of $Z_1\times Z_2$ ($Z_1$ and $Z_2$ are the charge
numbers of the colliding nuclei) are presented in
Fig.~\ref{14_fig}. One can see the exponential decrease
of the fusion probability with increasing $Z_1\times Z_2$
in the symmetric and almost symmetric reactions.
Therefore, the experimentally observed \cite{arm} rapid fall-off
of the fusion cross sections with increasing $Z_1\times Z_2$
is simply explained in our model.

The fusion and quasifission rates decrease with increasing
$\Gamma$. However, $P_{CN}$  increases quickly (the quasifission
rate decreases more strongly than the fusion rate) and reaches a
plateau at $\Gamma \approx 4$MeV because the changes of the
dissipative effects in $\eta$ and $R$ variables start to
compensate each other.  The calculated dependence of
$\lambda^{Kr}_\eta$, $\lambda^{Kr}_R$ and $P_{CN}$ on the
friction parameter $\Gamma$ is shown in Fig. ~\ref{2_fig} for
the $^{110}$Pd+$^{110}$Pd reaction.  Since the fusion
probability increases not much with $\Gamma$, our
calculations agree  with the values
of $P_{CN}$ obtained within the approach \cite{1} based on
statistical assumptions
within an order of magnitude.
It is seen that the results of the
calculations are not crucial to the exact value of the parameter
$\Gamma$ for
$\Gamma>2$ MeV.  A
realistic value of the parameter $\Gamma$ is about 2 MeV.  This
value should be used in the further calculations.

The dependences of $P_{CN}$ on the excitation energy
$E^*=E_{\rm c.m.}-V(R_m)$ of the DNS are given
for the $^{110}$Pd+$^{110}$Pd
and $^{86}$Kr+$^{136}$Xe reactions  in Fig.~\ref{3_fig}.  We
observe that $P_{CN}$ increases with $E^*$ because the value of
$\lambda^{Kr}_{R}$ in (\ref{2_eq}) increases slower than
$\lambda^{Kr}_{\eta}$ grows.  For the  $^{86}$Kr+$^{136}$Xe
reaction, the $P_{CN}$  values  were calculated for two cases.
In the first case, liquid drop masses in (\ref{7_eq}) and
spherical nuclei in the DNS were used to calculate $B^*_{fus}$
and $B_{qf}$. In the second case, realistic masses in
(\ref{7_eq}), spherical nuclei in the initial DNS and a deformed
heavy nucleus in the ground state near $\eta=\eta_{BG}$ were
taken. The pole orientation of the nuclei leads to the minimum
of the potential energy in the DNS.  Due to the deformation
effect near $\eta=\eta_{BG}$, the value of $V(R_m)$  decreases
as  compared to the calculation with the spherical nuclei and
$U(R_m,\eta_{BG})$ decreases. As a
result,  $B_{fus}^*$  decreases and $P_{CN}$  increases
(Fig.~\ref{3_fig}). The driving potential with
realistic binding energies and deformation effects
is preferable for small excitation energies. The driving potential
with the liquid drop binding energies is good for large
excitation energies. For the 5$n$ channel in the
$^{86}$Kr+$^{136}$Xe reaction, the excitation energy of the compound
nucleus is about 46 MeV ($E^*=30$ MeV) and $P_{CN}=4\times 10^{-2}$.
With $\sigma_c=23$ mb estimated with the model \cite{3} and
average value of $<\Gamma_n/\Gamma_f>=0.3$ taken from Ref.~\cite{24},
we obtain the evaporation residue cross section
$\sigma_{ER}\approx \sigma_c P_{CN} <\Gamma_n/\Gamma_f>^5=2.2\ \mu{\rm b}$
which is in agreement with the experimental value
$5\ \mu{\rm b}$ \cite{st}.

The energy threshold
for complete fusion is related to
the fusion barrier $B^*_{fus}$ (see Table~3) and can be much
smaller than the extra-extra push energy which,
for example, is $E_{\rm xx}$= 60 MeV and 30 MeV in
$^{110}$Pd+$^{110}$Pd and
$^{62}$Ni+$^{208}$Pb reactions, respectively,
predicted in the macroscopic dynamical model \cite{2}.
This result of our model is in agreement with recent experimental
data on the synthesis of the new superheavy elements \cite{5,25}.

\subsection{Small excitation energies of the DNS}
For low excitation energies of the initial DNS in asymmetrical
reactions $^{62}$Ni+$^{208}$Pb $\to ^{270}110$,
$^{70}$Zn+$^{208}$Pb $\to ^{278}112$, $^{82}$Se+$^{208}$Pb $\to
^{290}116$, and $^{48}$Ca+$^{244}$Pu $\to ^{292}114$ which were
used for the synthesis of new superheavy elements, the initial
DNS is in the local minimum of $U(\eta)$ due to shell effects
(realistic binding energies
are taken in (\ref{7_eq})) \cite{9}. In this
case, we can use the  Kramers--type expressions (\ref{5_eq})  to
estimate the values of $P_{CN}$ (Table~4).  For these
reactions, deformations of the DNS nuclei were taken into account
\cite{9,10}.  Since in the $^{48}$Ca+$^{244}$Pu reaction the heavy
nucleus is deformed even in the initial DNS, the treatment of
the deformation of nuclei leads to larger values of $B^*_{fus}$
as compared to the calculation with spherical nuclei
(Table~4). Deformation effects lead
to a decrease of $B^*_{fus}$ in  other reactions. The detailed
discussion of the influence of the deformation and orientation
of the DNS nuclei on the value of $B^*_{fus}$ is given in
\cite{10}. In the reactions leading to the superheavy nuclei
only partial waves with very small $J$ up to $J_{max}=10-15\hbar$
contribute the
evaporation residue cross section because
these nuclei are instable against fission for larger $J$.
The effect of the transient time in these reactions
seems to be very small.

Using the results presented in Table~4 one may explain the
smaller fusion yields of the nuclei with $Z=112$ as compared to
the fusion yields of the nuclei with $Z=110$ \cite{25}.  As
given in Table~4, the probability to obtain a superheavy nucleus
with $Z=116$ in the $^{82}$Se+$^{208}$Pb reaction is very small.
The use of this combination for producing heavy compound nuclei
with a small excitation energy  may be problematic.  In spite
of the larger value of $P_{CN}$ in the $^{48}$Ca+$^{244}$Pu
reaction as compared with others in Table~4, the compound
nucleus seems to be more excited due to the $Q$--value.  The
analysis of the surviving probability $W_{sur}$ of the compound
nucleus is extremely important to estimate the yield of the
element with $Z=114$ in this reaction.

By comparing  $P_{CN}$ in Tables~3 and 4 with $P_{CN}$ given in
the experimental work, we should bear in mind that the
experimental values  were extracted from the evaporation
residues cross sections $\sigma_{ER}(E_{\rm c.m.})$  by model
assumptions about the surviving probability $W_{sur}$ of the
excited compound nucleus
($\sigma_{ER}= \sigma_c\,P_{CN}\,W_{sur}$) \cite{1,22}.
For the 1n channel in
the $^{62}$Ni+$^{208}$Pb reaction, the excitation energy of the
compound nucleus is about 13 MeV and $P_{CN}=7\times10^{-6}$.
Extrapolating the systematic representation in
Refs.~\cite{24,26} for the nucleus $^{270}110$ we estimated
$W_{sur}\approx \Gamma_n/\Gamma_f=3\times 10^{-4}$.  With
$\sigma_c=4$ mb estimated with the optical model \cite{3} and
the values of $P_{CN}$ and $W_{sur}$, we obtain $\sigma_{ER}= 8.4$
pb which is in agreement with the experiment \cite{5}.

A discussion on the accuracy of the model presented is
necessary for applying it to reactions
producing superheavy elements with very small cross
sections. As in any model, certain assumptions are used in our
approach. However, with the same assumptions and the set of the
parameters our model is able to describe the experimental data
for different reactions. As it is seen, our model gives good
results for the $^{90}$Zr+$^{90}$Zr reaction for which the known
traditional models are applicable. However, our model describes
also the experimental data well in the case of  reactions where
the fusion cross sections are very small
and other models may fail \cite{1}.

\section{Summary}
The new model suggested to calculate the probability of the
fusion of heavy nuclei is useful for the analysis of
experimental data.  The results obtained support the use of the
simple statistical assumptions of our  previous studies
\cite{1,9}.  A good description of $P_{CN}$ in our model can be
considered as an evidence for the DNS--concept providing a
realistic interpretation of the mechanism of fusion process.  In
our opinion the fusion probability is small in many reactions
with heavy nuclei because quasifission plays a major role.
Without taking  the quasifission  into account, the explanation of the
experiments on the fusion of heavy nuclei is not possible.
Based on the results presented, we plan
calculations of the evaporation residues cross sections for a
large set of reactions used to produce the superheavy
elements.

\acknowledgments
We thank Dr. E.Cherepanov, Dr. A.Nasirov (Dubna) and
Dr. S.Hofmann (Darmstadt) for fruitful discussions.
The authors
(G.G.A. and N.V.A.) are grateful to the
Justus--Liebig--Universit\"at Giessen for the hospitality and
financial support. This work was supported in part by
the Russian Foundation for Basic Research under Grants N
95--02--05684, N 95--02--05975, and DFG.
One of the authors (N.V.A.) thank the Alexander von Humboldt
Foundation for the support during the completion this work.

\eject
\begin{table}
\caption{Calculated transient times (\protect\ref{91_eq}) and
(\protect\ref{92_eq}), quasistationary values of $1/\lambda_i^{Kr}$
and life time (\protect\ref{24_eq})
in the symmetric and almost symmetric
reactions for $J=0$, $\Gamma=2$ MeV and
the excitation energy $E^{*}=30$ MeV of
the initial DNS.
The calculations were made with
$B^*_{fus}$ and $B_{qf}$ given in Table~3.
}
\label{11_tab}
\begin{tabular}{|c|c|c|c|c|c|c|}
Reactions & $\tau_R$ & $1/\lambda_R^{Kr}$ & $\tau_\eta$ &
$1/\lambda_\eta^{Kr}$ &
$t_{00}$ & $\Delta t$ \\
 & $\times 10^{-21}$s & $\times 10^{-21}$s &$\times 10^{-21}$s &
 $\times 10^{-19}$s  & $\times 10^{-21}$s  & $\times 10^{-21}$s  \\
\hline
$^{90}$Zr+$^{90}$Zr & 1.2 & 104 & 2.4 &
1.6 & 63 & 1.0 \\ \hline
$^{100}$Mo+$^{100}$Mo & 1.1 & 60 & 2.8 & 39 &
58 & 0.6 \\ \hline
$^{110}$Pd+$^{110}$Pd & 1.0 & 32 & 3.1 &
2800 & 32 & 0.6  \\ \hline
$^{86}$Kr+$^{136}$Xe & 1.1 & 70 & 2.8 &
18 & 68 & 0.6 \\ \hline
$^{110}$Pd+$^{136}$Xe & 0.5 & 50 & 3.2 &
8300& 5 & 0.3 \\ \hline
$^{136}$Xe+$^{136}$Xe & 0.5 & 50 & 3.5 &
7.1$\times 10^{6}$& 5 & 0.3
\end{tabular}
\end{table}

\begin{table}
\caption{Calculated values of $P_{CN}^0$, $\Delta P_{CN}$
and the ratio $\Delta P_{CN}/P_{CN}^0$
(see Eq.~(\protect\ref{24_eq}))
in the symmetric and almost symmetric
reactions for $J=0$, $\Gamma=2$ MeV and
the excitation energy $E^{*}=30$ MeV of
the initial DNS.
The calculations were made with
$B^*_{fus}$ and $B_{qf}$ given in Table~3.
}
\label{12_tab}
\begin{tabular}{|c|c|c|c|}
Reactions & $P_{CN}^0$ & $\Delta P_{CN}$ & $\Delta P_{CN}/P_{CN}^0$
 \\ \hline
$^{90}$Zr+$^{90}$Zr & $4.0\times 10^{-1}$ & $2.7\times 10^{-3}$ &
$6.7\times 10^{-3}$ \\
$^{100}$Mo+$^{100}$Mo & $1.5\times 10^{-2}$ & $2.6\times 10^{-4}$ &
$1.7\times 10^{-2}$ \\
$^{110}$Pd+$^{110}$Pd & $1.1\times 10^{-4}$ & $4.2\times 10^{-6}$ &
$3.8\times 10^{-2}$ \\
$^{86}$Kr+$^{136}$Xe & $4.0\times 10^{-2}$ & $5.6\times 10^{-4}$ &
$1.4\times 10^{-2}$  \\
$^{110}$Pd+$^{136}$Xe & $5.5\times 10^{-6}$ & $1.7\times 10^{-6}$ &
$3.1\times 10^{-1}$  \\
$^{136}$Xe+$^{136}$Xe & $6.1\times 10^{-9}$ & $1.9\times 10^{-9}$ &
$3.1\times 10^{-1}$
\end{tabular}
\end{table}

\begin{table}
\caption{Calculated fusion  probability
$P_{CN}$
in the symmetric and almost symmetric
reactions for different friction parameters $\Gamma$.
The calculations were made for $J=0$
and the excitation energy $E^{*}=30$ MeV of
the initial DNS.
For the calculated values of $P_{CN}$, liquid drop masses
and spherical nuclei in the DNS
were used in (\protect\ref{7_eq}). }
\label{1_tab}
\begin{tabular}{|c|c|c|c|c|c|c|c|}
Reactions & $B_{fus}^*$ & $B_{qf}$ & \multicolumn{5}{c|}{$P_{CN}$}
\\ \cline{4-8}
 &(MeV) &(MeV) &  &  &   &  &  \\
& & & $\Gamma$=0 MeV & $\Gamma$=1 MeV & $\Gamma$=2 MeV &
$\Gamma$=3 MeV & $\Gamma$=4 MeV \\\hline
$^{90}$Zr+$^{90}$Zr & 6 & 5 & 3.1$\times 10^{-1}$ &
3.6$\times 10^{-1}$ &
4.0$\times 10^{-1}$ & 4.5$\times 10^{-1}$ &
4.8$\times 10^{-1}$ \\ \hline
$^{100}$Mo+$^{100}$Mo & 10 & 4 & 8.3$\times 10^{-3}$ &
1.0$\times 10^{-2}$ &
1.5$\times 10^{-2}$ & 1.8$\times 10^{-2}$ &
2.0$\times 10^{-2}$ \\ \hline
$^{110}$Pd+$^{110}$Pd & 15 & 3 & 6.5$\times 10^{-5}$ &
7.9$\times 10^{-5}$ &
1.1$\times 10^{-4}$ & 1.4$\times 10^{-4}$ &
1.5$\times 10^{-4}$  \\ \hline
$^{86}$Kr+$^{136}$Xe & 8.5 & 4 & 2.3$\times10^{-2}$ &
2.7$\times10^{-2}$ &
4.0$\times10^{-2}$ & 5.0$\times10^{-2}$ &
5.6$\times10^{-2}$ \\ \hline
$^{110}$Pd+$^{136}$Xe & 15.5 & 0.5 & 9.0$\times 10^{-7}$ &
2.2$\times 10^{-6}$&
3.8$\times 10^{-6}$ & 4.4$\times 10^{-6}$ &
5.0$\times 10^{-6}$ \\ \hline
$^{136}$Xe+$^{136}$Xe & 22.5 & 0.5 & 1.6$\times 10^{-9}$ &
2.5$\times 10^{-9}$&
4.2$\times 10^{-9}$ & 4.9$\times 10^{-9}$ &
5.7$\times 10^{-9}$
\end{tabular}
\end{table}

\begin{table}
\caption{The same as in Table~3, but for
asymmetric reactions used for producing superheavy nuclei.
The deformation effects
are taken into account (see text).
The calculations were made for $J=0$ and the excitation energy $E^{*}=15$
MeV of the initial DNS with exception of the
$^{48}$Ca+$^{244}$Pu  reaction.
For the $^{48}$Ca+$^{244}$Pu reaction, the
choice of the initial  excitation energies in the calculations of
$P_{CN}$ with ($E^*=33$ MeV) and without
($E^*=15$ MeV) deformation effects
(sph.) yields the same excitation energy of
the compound nucleus ($\approx 40$ MeV).}
\label{2_tab}
\begin{tabular}{|c|c|c|c|c|c|c|c|}
Reactions & $B_{fus}^*$ & $B_{qf}$ &
\multicolumn{5}{c|}{$P_{CN}$}\\ \cline{4-8}
 &(MeV) &(MeV) &  &  &   &  &  \\
 & & & $\Gamma$=0 MeV & $\Gamma$=1 MeV & $\Gamma$=2 MeV &
 $\Gamma$=3 MeV & $\Gamma$=4 MeV \\\hline
$^{62}$Ni+$^{208}$Pb $\to ^{270}110$ & 8 & 1.5 &
1.4$\times 10^{-4}$  &
1.6$\times 10^{-4}$ & 2.2$\times 10^{-4}$ &
2.9$\times 10^{-4}$ &
3.2$\times 10^{-4}$ \\ \hline
$^{70}$Zn+$^{208}$Pb $\to ^{278}112$ & 9.5 & 1 &
9.1$\times 10^{-6}$  &
1.1$\times 10^{-5}$ & 1.5$\times 10^{-5}$ &
1.9$\times 10^{-5}$ &
2.2$\times 10^{-5}$ \\ \hline
$^{82}$Se+$^{208}$Pb $\to ^{290}116$ & 12.5 & 0.5 &
9.4$\times 10^{-8}$ &
1.1$\times 10^{-7}$ & 1.5$\times 10^{-7}$ &
1.9$\times 10^{-7}$ &
2.2$\times 10^{-7}$ \\ \hline
$^{48}$Ca+$^{244}$Pu $\to ^{292}114$ & 12 & 4 &
3.7$\times 10^{-4}$ &
4.4$\times 10^{-4}$ & 6.2$\times 10^{-4}$ &
8.0$\times 10^{-4}$ &
9.1$\times 10^{-4}$ \\ \hline
(sph.) $^{48}$Ca+$^{244}$Pu $\to ^{292}114$  & 7 & 3 &
2.3$\times 10^{-3}$ &
2.8$\times 10^{-3}$ & 3.9$\times 10^{-3}$ & 5.0$\times 10^{-3}$ &
5.6$\times 10^{-3}$
\end{tabular}
\end{table}


\begin{figure}
\caption{  Calculated potential energy of the DNS
corresponding to the $^{180}$Hg compound nucleus
as a function of the mass asymmetry for zero angular momentum
(upper part).
The energy scales are
normalized to the total energy of the compound nucleus.
Liquid drop binding energies are used in (\protect\ref{7_eq}).
Calculated nucleus-nucleus potential  in the
$^{90}$Zr+$^{90}$Zr
reaction for zero angular momentum (bottom part).}
\label{1_fig}
\end{figure}

\begin{figure}
\caption{Calculated potential energy of the DNS
in the $^{86}$Kr+$^{136}$Xe reactions as a function
of $\eta$ for $J=0$. The result calculated with
liquid drop binding energies in (\protect\ref{7_eq})
is presented by the dashed line.
The driving potentials with and without deformation effects
in the DNS, obtained with realistic binding energies
in (\protect\ref{7_eq}), are presented by thick and thin lines,
respectively.
The deformed nuclei of the DNS are assumed in
the pole orientation.}
\label{11_fig}
\end{figure}

\begin{figure}
\caption{Dependences of $B_{qf}$, $B^*_{fus}$, $P_{CN}$ and
$(2J+1)P_{CN}$ on $J$ for the reactions $^{90}$Zr+$^{90}$Zr
(solid lines) and  $^{110}$Pd+$^{110}$Pd (dashed lines).
In the calculations of $P_{CN}$ the values
$E^*=30$ MeV and $\Gamma=2$ MeV are used.}
\label{12_fig}
\end{figure}

\begin{figure}
\caption{Calculated fusion rates $\lambda_\eta$ as a
function of time for the reactions
$^{90}$Zr+$^{90}$Zr (solid line) and $^{136}$Xe+$^{136}$Xe
(dashed line) at
$E^*=30$ MeV and $\Gamma=2$ MeV.}
\label{13_fig}
\end{figure}

\begin{figure}
\caption{Calculated values (solid points) of the
fusion probability as a function of $Z_1\times Z_2$
for the reactions presented in Table~3 and the reactions
$^{96}$Zr+$^{124}$Sn ($B^*_{fus}=9$ MeV, $B_{qf}=4$ MeV)
and $^{124}$Sn+$^{124}$Sn ($B^*_{fus}=16$ MeV, $B_{qf}=0.5$ MeV)
at $J=0$, $\Gamma=2$ MeV and $E^*=30$ MeV.
$Z_1$ and $Z_2$ are the charge numbers of the colliding nuclei.
The solid line is drawn to guide the eye.}
\label{14_fig}
\end{figure}

\begin{figure}
\caption{
Fusion (long-dashed line) and
quasifission (short-dashed line) rates
$\lambda^{Kr}_i (i=\eta,R)$ and fusion  probability
 $P_{CN}$ (solid line) as a function of the
friction parameter $\Gamma$ for the
$^{110}$Pd+$^{110}$Pd reaction at $J=0$.}
\label{2_fig}
\end{figure}

\begin{figure}
\caption{Fusion probability $P_{CN}$ as a function of the
excitation energy
$E^*=E_{\rm c.m.}-V(R_m)$ of the initial DNS  for the reactions
$^{110}$Pd+$^{110}$Pd (upper part) and
$^{86}$Kr+$^{136}$Xe (bottom part) at $J=0$.
The results were obtained by using liquid-drop masses and
spherical nuclei (solid lines)
and realistic masses and a deformed heavy nucleus
(dashed line) in the DNS.
The calculations were made with a
friction parameter $\Gamma=2$ MeV.}
\label{3_fig}
\end{figure}

\newpage
\epsfig{figure=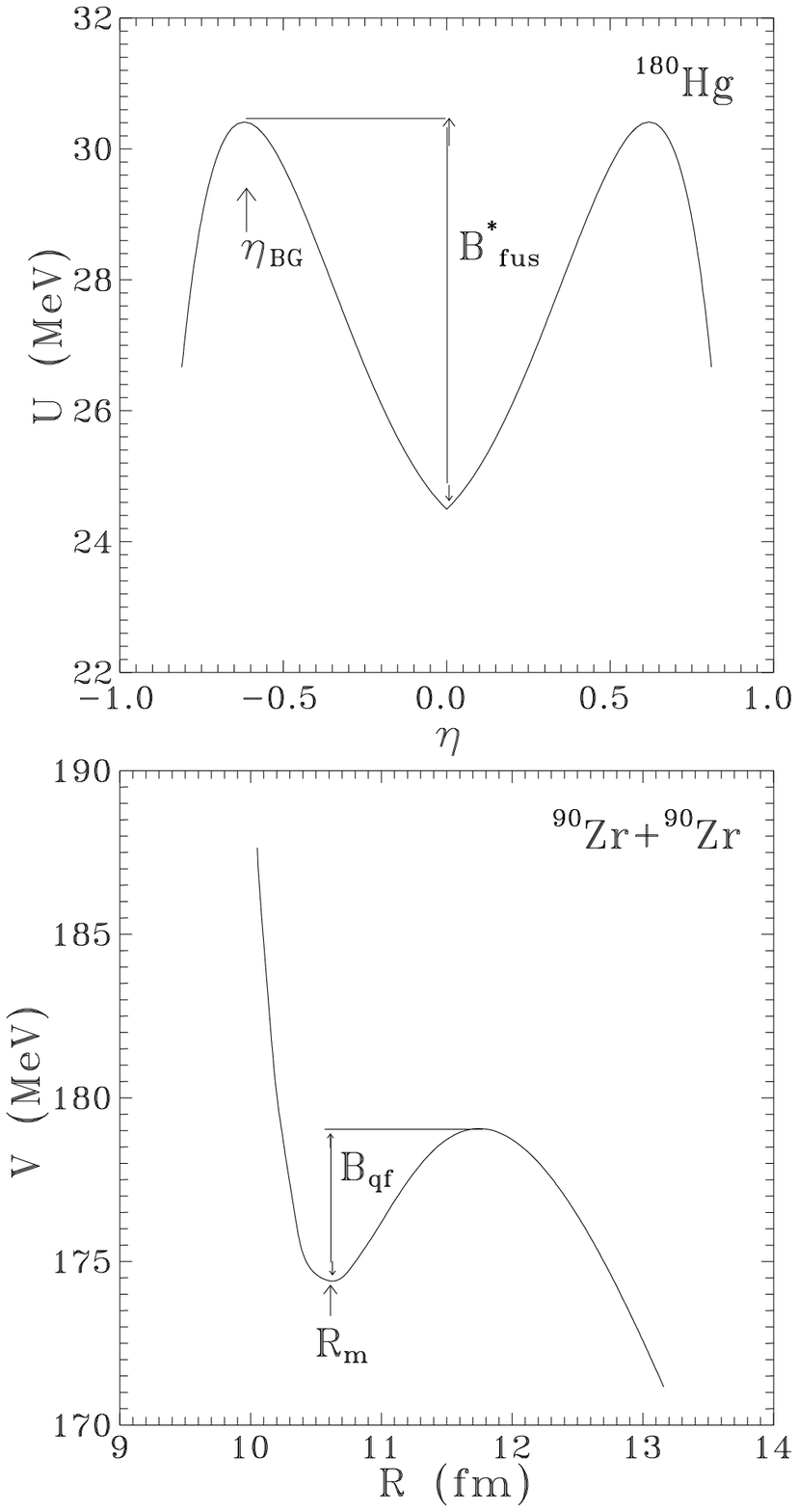,width=16cm,height=23cm}
\newpage
\epsfig{figure=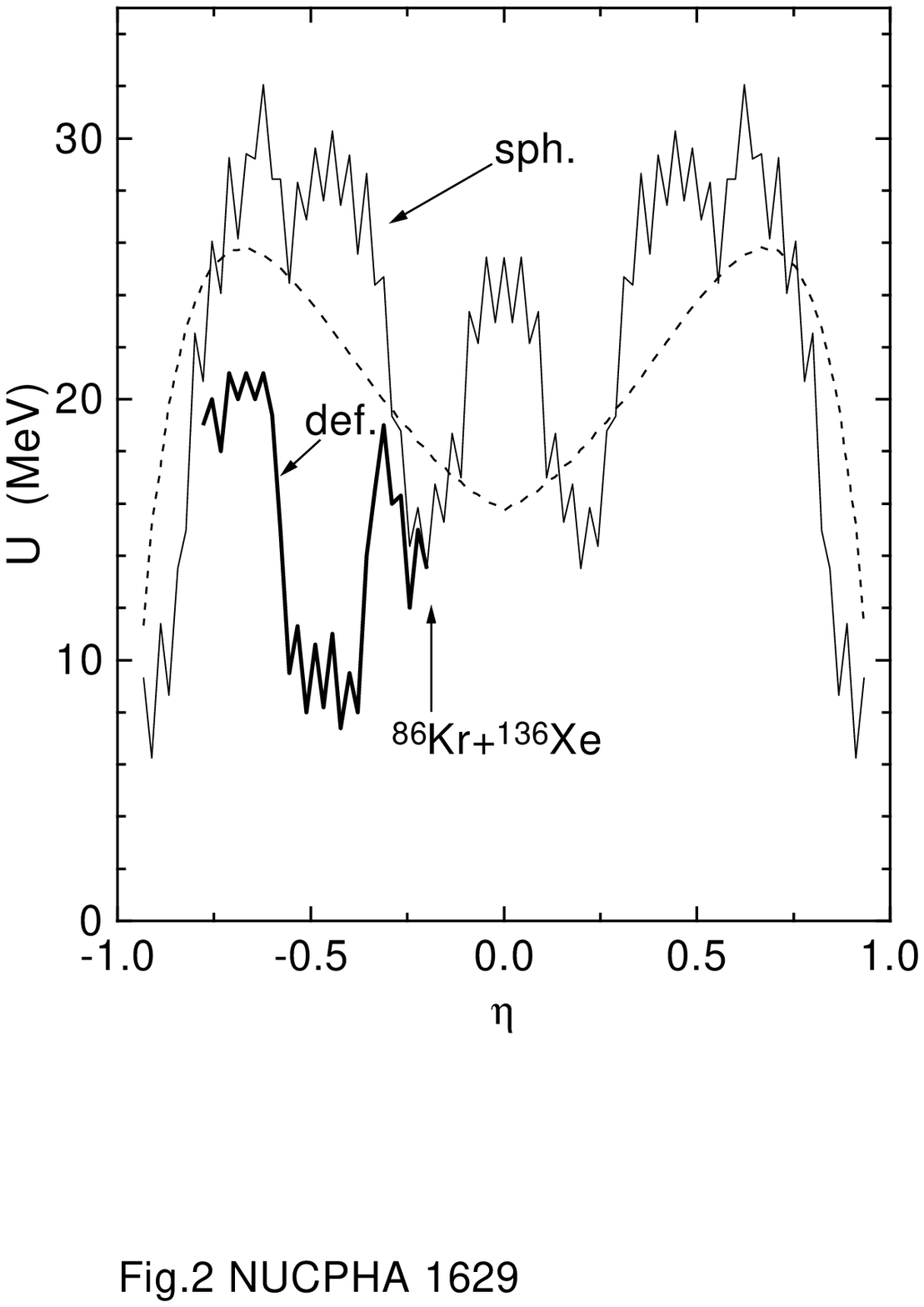,width=16cm,height=23cm}
\newpage
\epsfig{figure=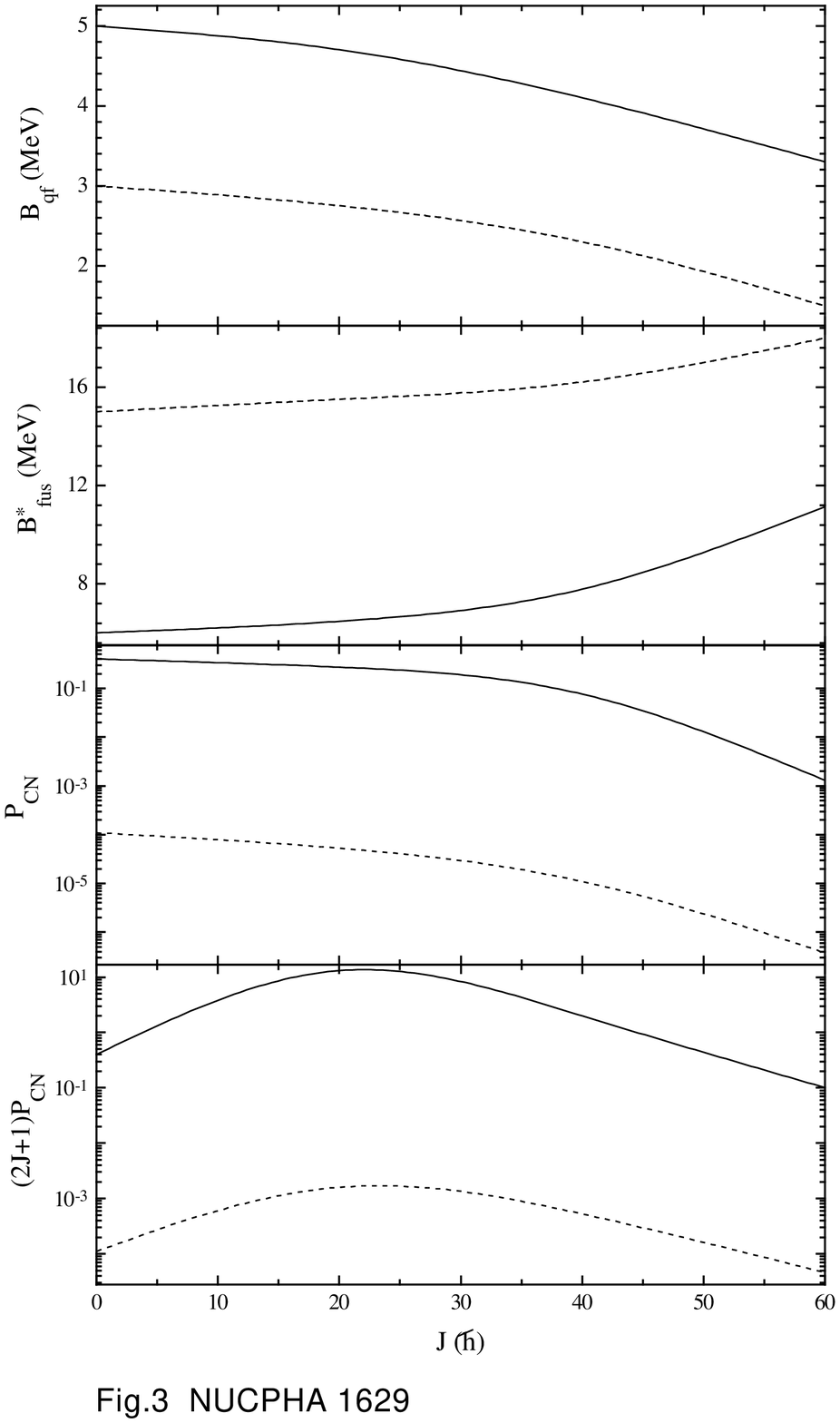,width=16cm,height=23cm}
\newpage
\epsfig{figure=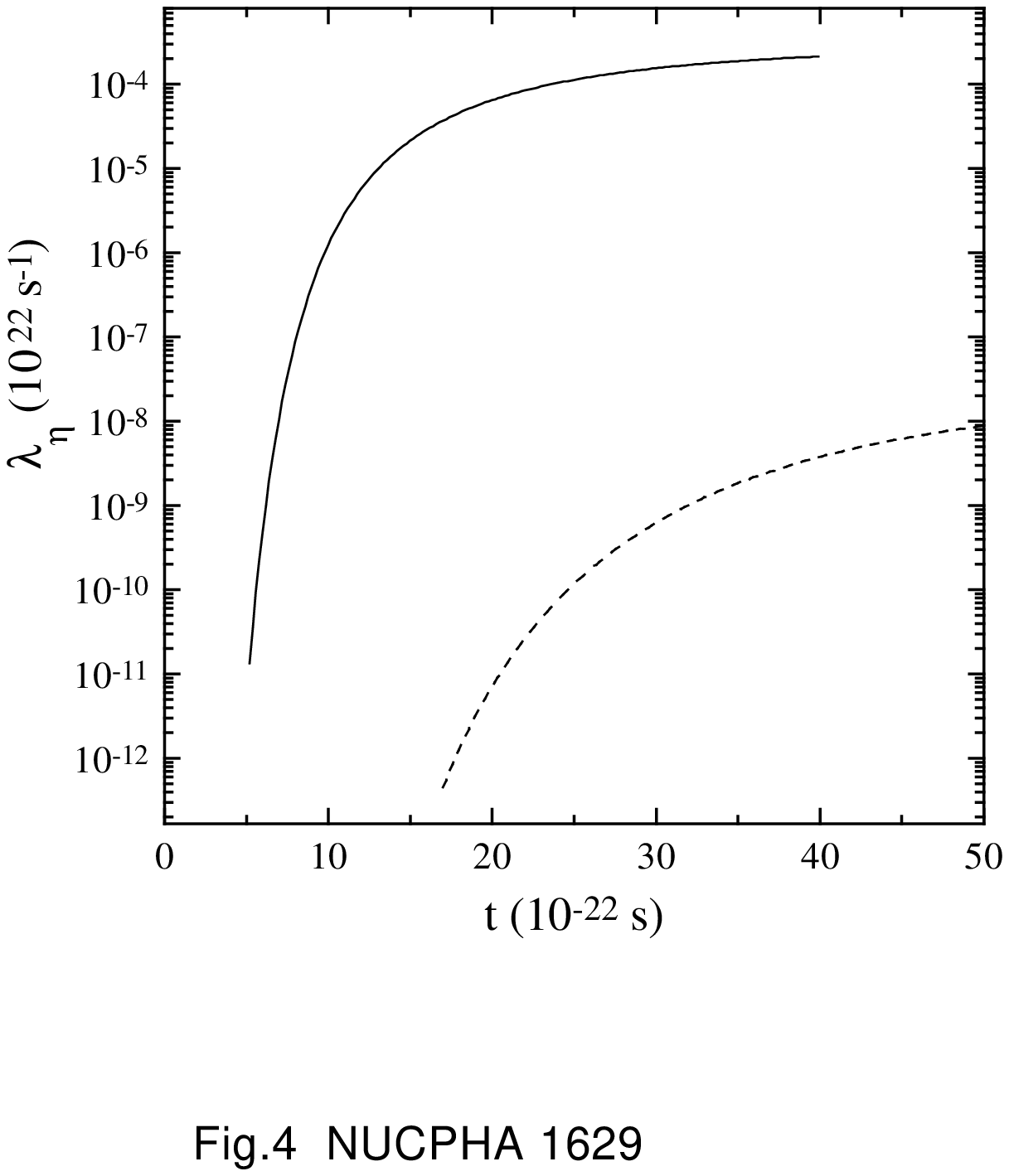,width=16cm,height=23cm}
\newpage
\epsfig{figure=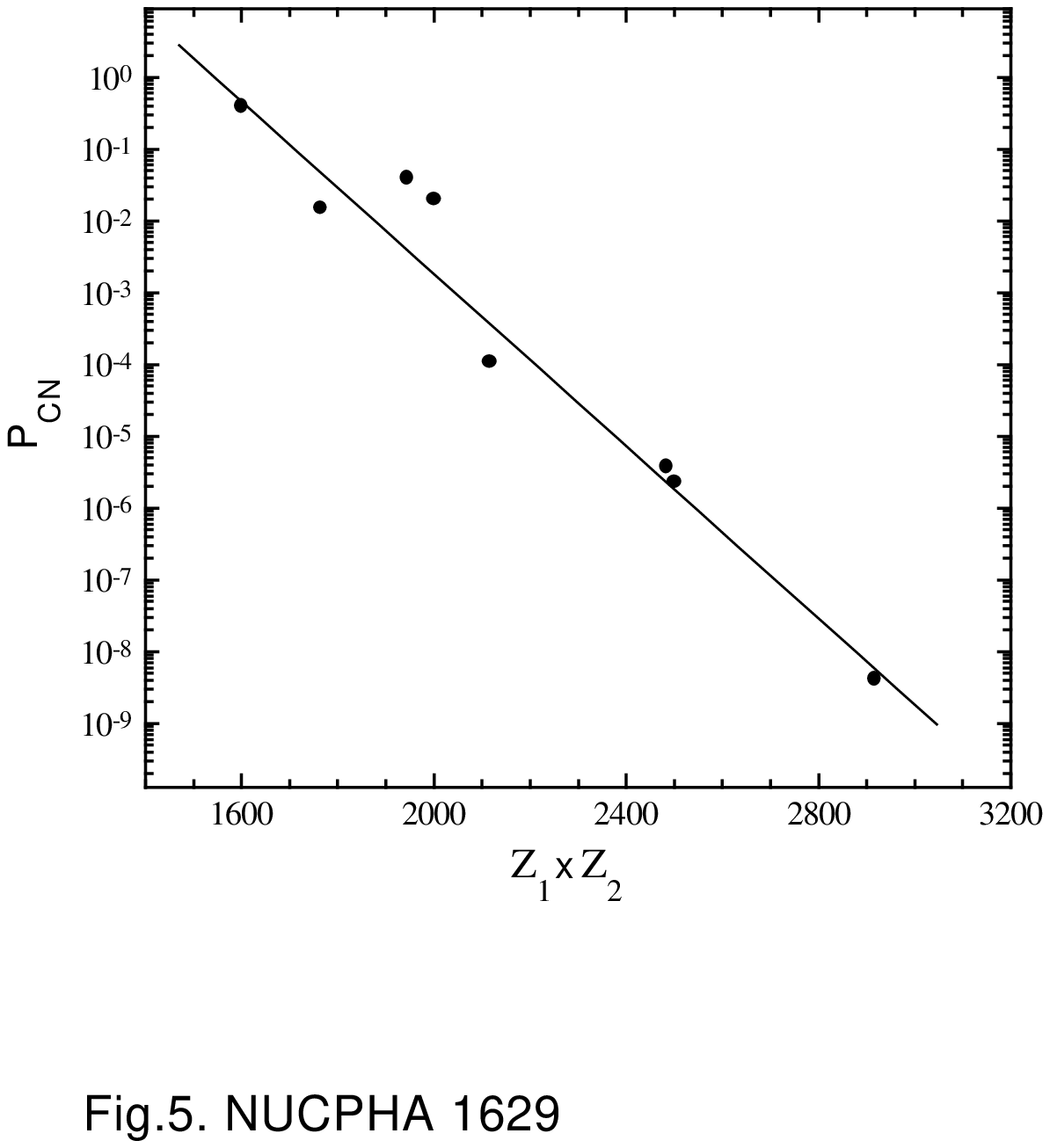,width=16cm,height=23cm}
\newpage
\epsfig{figure=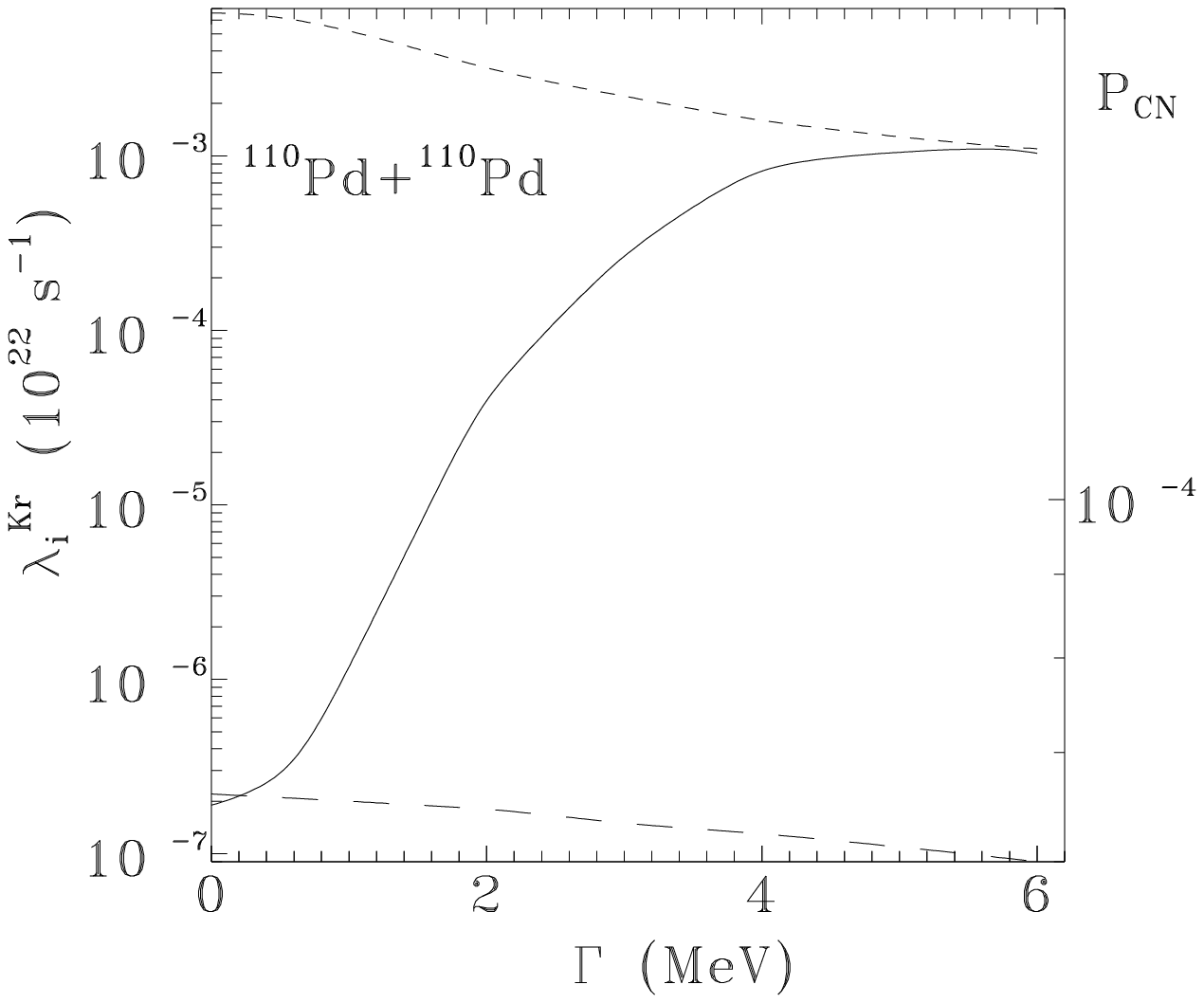,width=15cm,height=12cm}
\newpage
\epsfig{figure=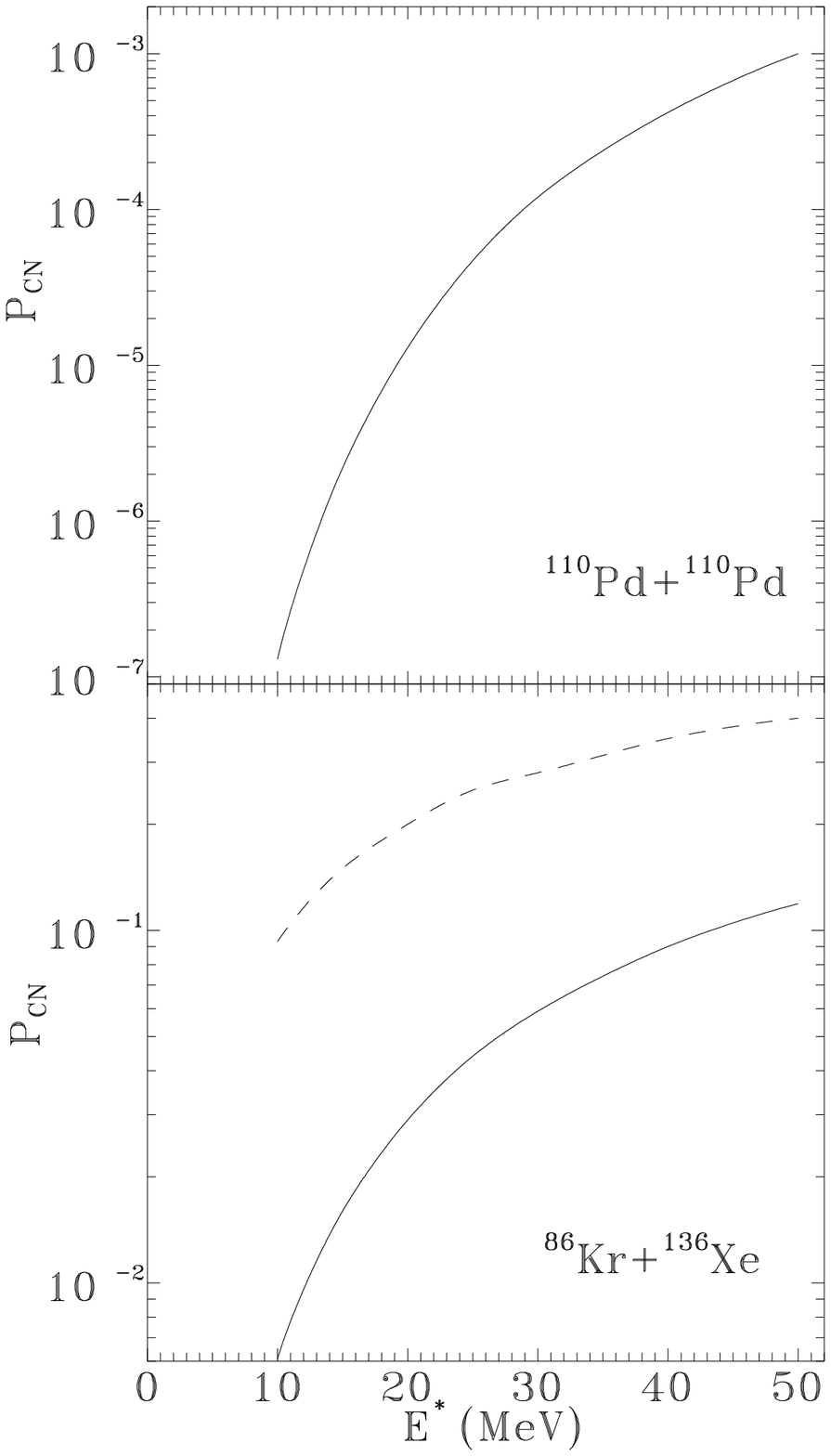,width=14cm,height=22cm}


\begin{thebibliography}{99}
\bibitem{1} N.V.Antonenko, E.A.Cherepanov, A.V.Nasirov,
V.B.Permjakov and
V.V.Volkov, Phys. Lett. B 319 (1993) 425;
Phys. Rev. C 51 (1995) 2635;
E.A.Cherepanov, V.V.Volkov, N.V.Antonenko, V.B.Permjakov
and A.V.Nasirov, Nucl. Phys. A 583 (1995) 165.
\bibitem{2} W.J. Swiatecki, Prog. Particle and Nucl. Phys.
{4} (1980) 383; Phys. Scripta {24} (1981) 113;
S.Bjornholm and W.J.Swiatecki, Nucl. Phys. A 391 (1982) 471;
J.P.Blocki, H.Feldmeier and W.J.Swiatecki,
Nucl. Phys. A 459 (1986) 145.
\bibitem{3} A.S. Iljinov, Yu.Ts. Oganessian and E.A. Cherepanov,
Sov. J. Nucl. Phys. {36} (1982) 118.
\bibitem{4} D.H.E. Gross, R.C. Nayak and
L. Satpathy, Z. Phys.  A { 299} (1981) 63;
P. Fr\"{o}brich,  Phys. Rep. {116} (1984) 337;
J. Marten and P. Fr\"{o}brich, Nucl. Phys A 545 (1992) 854.
\bibitem{5} S. Hofmann, V. Ninov, F.P. Hessberger, P. Armbruster,
H. Folger, G. M\"unzenberg, H.J. Schott, A.G. Popeko,
A.V. Eremin, A.N. Andreev,
S. Saro, R. Janic and M. Leino, Z. Phys. A 350 (1995) 277;
Z. Phys.  A 350 (1995) 281.
\bibitem{25} S. Hofmann et al., Z. Phys. A 354 (1996) 229.
\bibitem{6} W.Morawek, T.Ackermann, T.Brohn, H.G.Clerc,
U.Gollerthan, E.Hanlet,
M.Horz, W.Schwab, B.Voss, K.H.Schmidt,
J.J.Gaimard, F.P.Hessberger,
in: GSI Scientific Report (GSI, 1988) p.38;
K.H.Schmidt, W.Morawek, Rep. Prog. Phys. 54 (1991) 949;
W.Morawek, T.Ackermann, T.Brohn, H.G.Clerc, U.Gollerthan, E.Hanlet,
M.Horz, W.Schwab, B.Voss, K.H.Schmidt and F.P.Hessberger,
Z. Phys. A 341 (1991) 75.
\bibitem{arm} P.Armbruster, Ann. Rev. Nucl. Part. Sci. 35 (1985) 135.
\bibitem{7} V.V.Volkov, Proc. Intern. School-Seminar on Heavy
Ion Physics (Dubna, 1986), D7-87-68 (JINR, Dubna, 1987) p.528;
Izv. AN SSSR ser. fiz. {50} (1986) 1879;
in: Proc. of the 6th Intern. Conf. on
Nuclear Reaction Mechanisms
(Varenna, Italy 1991), ed. E.Gadioli (Ricerca Scientifica ed
Educazione Permanente Supplemento n. 84, 1991) p.39.
\bibitem{8} V.V. Volkov,
Nuclear reactions of deep inelastic transfers
(Energoizdat, Moscow, 1982);
W.U. Schr\"oder and  J.R. Huizenga, {\it in}
Treatise~on~Heavy-Ion~Science, ed. D.A. Bromley,
v.2  (New York, Plenum Press, 1984) p. 115.
\bibitem{9} V.V.Volkov, E.A.Cherepanov, N.V.Antonenko
and A.V.Nasirov,
in: Proc. Int. Conf. on Low Energy Nuclear Dynamics,
St.Petersburg, 1995, eds. Yu.Oganessian, R.Kalpakchieva,
W. von Oertzen (World Scientific, Singapore, 1995) p. 336.
\bibitem{10} G.G.Adamian, N.V.Antonenko
and W.Scheid,  Nucl. Phys.  A 618 (1997) 176.
\bibitem{111} K.H. Bhatt, P. Grange and B. Hiller,
Phys. Rev. C 33 (1986) 954;
P. Grange, Nucl. Phys. A 428 (1984) 37c.
\bibitem{11}  H.A. Kramers, Physica. 7 (1940) 284.
\bibitem{12}  V.M. Strutinsky, Phys. Lett. B 47 (1973) 121;
H. Hofmann and J.R. Nix, Phys. Lett. B 122 (1983) 117;
P. Grange, Jun-Qing Li and H.A. Weidenm\"uller,
Phys. Rev. C 27 (1983) 2063;
P. Fr\"obrich and G.R. Tillack, Nucl. Phys. A 540 (1992) 353.
\bibitem{13} H.A. Weidenm\"uller,  Jing-Shang Zhang,
J. Stat. Phys.  34 (1984) 191.
\bibitem{14} G.G.Adamian, N.V.Antonenko, R.V.Jolos
and A.K.Nasirov, Phys. Part. \& Nucl.  25 (1994) 583;
Nucl.  Phys.  A 551 (1993) 321;
N.V.Antonenko, S.P.Ivanova,
R.V.Jolos and W.Scheid, Phys. Rev. C 50 (1994) 2063.
\bibitem{15} H.J.Fink, J.Maruhn, W.Scheid and
W.Greiner, Z. Phys. A 268 (1974) 321;
J.Maruhn, W.Scheid and W.Greiner,
in: Heavy ion collisions, ed. R. Bock,
v. 2 (North-Holland, Amsterdam, 1980).
\bibitem{16} G.G.Adamian, N.V.Antonenko, R.V.Jolos
and W.Scheid,  Nucl. Phys. A (1997) in print.
\bibitem{17} I.I.Gonchar and G.I.Kosenko,
Sov. J. Nucl. Phys. 53 (1991) 133.
\bibitem{18}  H.Hofmann and P.J.Siemens,
Nucl. Phys. A 257 (1976) 165; R.Samhammer, H.Hofmann and
S.Yamaji, Nucl. Phys. A 503 (1989) 404.
\bibitem{19} G.G.Adamian, N.V.Antonenko and  R.V.Jolos,
Nucl. Phys. A 584 (1995) 205.
\bibitem{20}  A.M. Wapstra and G. Audi,
Nucl. Phys. A 440 (1985) 327.
\bibitem{21} G.G.Adamian, N.V.Antonenko,
R.V.Jolos, S.P.Ivanova and
O.I.Melnikova, Int. J. Mod. Phys. E 5 (1996) 191.
\bibitem{rom} S.Raman, C.H.Malarkey, W.T.Milner, C.W.Nestor
and P.H.Stelson, At. Data Nucl. Data Tables 36 (1987) 1.
\bibitem{ant} N.V.Antonenko, G.G.Adamain, V.V.Volkov, E.A.Cherepanov
and A.V.Nasirov, in: Proc. Int. Conf. Nuclear Structure at the Limits,
Argonne, 1996, (ANL, 1997) in print.
\bibitem{pom} K.Pomorski, J.Bartel, J.Richert and K.Dietrich
Nucl. Phys. A 605 (1996) 87.
\bibitem{22} P.Armbruster, in: Proc. Intern. School-Seminar
on Heavy Ion Physics (Dubna, 1986),
D7-87-68 (JINR, Dubna, 1987) p. 82.
\bibitem{23} H. Gaggeler, T. Sikkeland, G. Wirth,
W. Bruchle, W. Brugl, G. Franz, G. Herrmann,
J.V. Kratz, M. Schaedel, K. Summerer and W. Weber,
Z. Phys.  A 316 (1984) 291;
Nucl. Phys.  A 275 (1977) 464.
\bibitem{24} G.M\"unzenberg, Rep. Prog. Phys. 51 (1988) 57.
\bibitem{st} C.~Stodel, S.~Hofmann, F.P.Hessberger, V.Ninov,
R.N.Sagaidak, A.G.Popeko, Yu.Ts.Oganessian, A.Yu.Lavrentjev,
A.V.Eremin, in: GSI Scientific Report (GSI, 1996) p.17.
\bibitem{26} R.Vandenbosch and
J.R.Huizenga, Nuclear Fission (New York, Academic, 1973).

\end{thebibliography}
\end{document}